\documentclass[trackchanges]{aastex63}

\newcommand\aastex{AAS\TeX}
\newcommand\latex{La\TeX}
\newcommand{\ptfa}{PTF1\,J2238+7430}
\newcommand{\kms}{\ensuremath{{\rm km}\,{\rm s}^{-1}}}

\newcommand{\msol}{M$_\odot$}
\newcommand{\rsol}{R$_\odot$}
\newcommand{\teff}{T$_{\rm eff}$}
\newcommand{\logg}{$\log{g}$}
\newcommand{\porb}{$P_{\rm orb}$}
\newcommand{\vrot}{$v_{\rm rot}\sin{i}$}
\usepackage{multirow}
\usepackage{amssymb}
\usepackage{lineno}

\received{June 1, 2019}
\revised{January 10, 2019}
\accepted{\today}
\submitjournal{ApJL}

\shorttitle{A new double detonation progenitor system}
\shortauthors{Kupfer et al.}
\graphicspath{{./}{figures/}}

\begin{document}

\title{Discovery of a double detonation thermonuclear supernova progenitor}

\correspondingauthor{Thomas Kupfer}
\email{tkupfer@ttu.edu}

\author[0000-0002-6540-1484]{Thomas Kupfer}
\affil{Department of Physics and Astronomy, Texas Tech University, PO Box 41051, Lubbock, TX 79409, USA}

\author[0000-0002-4791-6724]{Evan B.~Bauer}
\affil{Center for Astrophysics $\mid$ Harvard \& Smithsonian, 60 Garden St, Cambridge, MA 02138, USA}

\author[0000-0002-2626-2872]{Jan van~Roestel}
\affiliation{Division of Physics, Mathematics and Astronomy, California Institute of Technology, Pasadena, CA 91125, USA}

\author[0000-0001-8018-5348]{Eric C. Bellm}
\affiliation{DIRAC Institute, Department of Astronomy, University of Washington, 3910 15th Avenue NE, Seattle, WA 98195, USA}

\author{Lars Bildsten}
\affiliation{Kavli Institute for Theoretical Physics, University of California, Santa Barbara, CA 93106, USA}
\affiliation{Department of Physics, University of California, Santa Barbara, CA 93106, USA}

\author{Jim Fuller}
\affiliation{Division of Physics, Mathematics and Astronomy, California Institute of Technology, Pasadena, CA 91125, USA}

\author{Thomas A. Prince}
\affiliation{Division of Physics, Mathematics and Astronomy, California Institute of Technology, Pasadena, CA 91125, USA}

\author[0000-0001-7798-6769]{Ulrich Heber}
\affiliation{Dr.\ Karl Remeis-Observatory \& ECAP, Astronomical Institute, Friedrich-Alexander University Erlangen-Nuremberg (FAU), Sternwartstr.\ 7, 96049 Bamberg, Germany}

\author{Stephan Geier}
\affiliation{Institut f\"ur Physik und Astronomie, Universit\"at Potsdam, Haus 28, Karl-Liebknecht-Str. 24/25, D-14476 Potsdam-Golm, Germany}

\author{Matthew J. Green}
\affiliation{Department of Astrophysics, School of Physics and Astronomy, Tel Aviv University, Tel Aviv 6997801, Israel}

\author[0000-0001-5390-8563]{Shrinivas R. Kulkarni}
\affiliation{Division of Physics, Mathematics and Astronomy, California Institute of Technology, Pasadena, CA 91125, USA}

\author{Steven Bloemen}
\affiliation{Department of Astrophysics/IMAPP, Radboud University Nijmegen, P.O. Box 9010, 6500 GL Nijmegen, The Netherlands}

\author[0000-0003-2451-5482]{Russ R. Laher}
\affiliation{IPAC, California Institute of Technology, 1200 E. California Blvd, Pasadena, CA 91125, USA}

\author[0000-0001-7648-4142]{Ben Rusholme}
\affiliation{IPAC, California Institute of Technology, 1200 E. California Blvd, Pasadena, CA 91125, USA}

\author{David Schneider}
\affiliation{Dr.\ Karl Remeis-Observatory \& ECAP, Astronomical Institute, Friedrich-Alexander University Erlangen-Nuremberg (FAU), Sternwartstr.\ 7, 96049 Bamberg, Germany}



\begin{abstract}

We present the discovery of a new double detonation progenitor system consisting of a hot subdwarf B (sdB) binary with a white dwarf companion with an \porb=76.34179(2)\,min orbital period. Spectroscopic observations are consistent with an sdB star during helium core burning residing on the extreme horizontal branch. Chimera light curves are dominated by ellipsoidal deformation of the sdB star and a weak eclipse of the companion white dwarf. Combining spectroscopic and light curve fits we find a low mass sdB star, $M_{\rm sdB}=0.383\pm0.028$\,\msol\, with a massive white dwarf companion, $M_{\rm WD}=0.725\pm0.026$\,\msol. From the eclipses we find a blackbody temperature for the white dwarf of $26,800$\,K resulting in a cooling age of $\approx$25\,Myrs whereas our \texttt{MESA} model predicts an sdB age of $\approx$170\,Myrs. We conclude that the sdB formed first through stable mass transfer followed by a common envelope which led to the formation of the white dwarf companion $\approx$25\,Myrs ago.

Using the \texttt{MESA} stellar evolutionary code we find that the sdB star will start mass transfer in $\approx$6\,Myrs and in $\approx$60\,Myrs the white dwarf will reach a total mass of $0.92$\,\msol\, with a thick helium layer of $0.17$\,\msol. This will lead to a detonation that will likely destroy the white dwarf in a peculiar thermonuclear supernova.  \ptfa\, is only the second confirmed candidate for a double detonation thermonuclear supernova. Using both systems we estimate that at least $\approx$1\,\% of white dwarf thermonuclear supernovae originate from sdB+WD binaries with thick helium layers, consistent with the small number of observed peculiar thermonuclear explosions.

\end{abstract}



\section{Introduction} \label{sec:intro}
Most hot subdwarf B stars (sdBs) are core helium burning stars with masses around 0.5 \msol\, and thin hydrogen envelopes (\citealt{heb86,heb09,heb16}). A large number of sdB stars are in close orbits with orbital periods of \porb$<10$\,days \citep{nap04a,max01}, with the most compact systems reaching orbital periods of $\lesssim 1$\,hour (e.g. \citealt{ven12,gei13,kup17,kup17a, kup20, kup20a}). The only way to form such tight binaries is orbital shrinkage through a common envelope phase followed by the loss of angular momentum due to the radiation of gravitational waves \citep{han02,han03,nel10a}.

SdB binaries with white dwarf (WD) companions which exit the common envelope phase at \porb$\lesssim$2\,hours will reach contact while the sdB is still burning helium \citep{bau21}. Due to the emission of gravitational waves the orbit of the binary will shrink until the sdB fills its Roche Lobe at a period of $\approx30-100$\,min, depending on the evolutionary stage and envelope thickness of the hot subwarf (e.g. \citealt{sav86,tut89,tut90,ibe91,yun08,pie14,bro15,neu19,bau21}).

The known population of sdB + WD binaries consists mostly of systems with orbital periods too large to start accretion before the sdB turns into a WD \citep{kup15a}. Currently only four detached systems with a WD companion are known to have \porb$<2$\,hours \citep{ven12,gei13,kup17,kup17a, pel21}. Just recently \citet{kup20,kup20a} discovered the first two Roche lobe filling hot subdwarfs as part of a high-cadence Galactic Plane survey using the Zwicky Transient Facility \citep{kup21}. Both systems can be best explained as Roche Lobe filling sdOB stars which have started mass transfer to a WD companion. The light curves in both systems show deep eclipses from an accretion disk. Due to their high effective temperatures, both sdOB stars are predicted to be in a short lived phase where the sdOB undergoes residual hydrogen shell burning. 

The most compact known sdB binary where the sdB is still undergoing core-helium burning is CD--30$^{\circ}$11223. The binary has an orbital period \porb=70.5\,min and a high mass WD companion ($M_{\rm WD}\approx0.75$\,\msol; \citealt{ven12,gei13}). The sdB in CD--30$^{\circ}$11223 will begin transferring helium to its WD companion in $\approx40$\,Myr when the system has shrunk to an orbital period \porb$\approx$40\,min. After the WD accretes $\approx$0.1\,\msol, helium burning is predicted to be ignited unstably in the accreted helium layer on the WD surface \citep{bro15,bau17}. This could either disrupt the WD even when the mass is significantly below the Chandrasekhar mass, a so-called double detonation supernova  (e.g. \citealt{liv90,liv95,fin10,woo11,wan12,she14,wan18}) or just detonate the He-shell without disrupting the WD which results in a faint and fast .Ia supernova with subsequent weaker He-flashes \citep{bil07,bro15}. Therefore, systems like CD--30$^{\circ}$11223 are predicted to be either the progenitors for double detonation thermonuclear supernovae or perhaps faint and fast .Ia supernovae that do not disrupt the WD. 

\citet{de19,de20} presented the discovery of a sample of calcium-rich transients consistent with a thick helium shell double detonation on a sub-Chandrasekhar-mass WD \citep{pol19,pol21}. The majority of these transients are located in old stellar populations with only a small sub-sample found in in star forming environments. 

The question remains just how common systems like CD--30$^{\circ}$11223 are. To address this question we have conducted a search for (ultra-)compact post-common envelope systems using the Palomar Transient Factory (PTF; \citealt{law09,rau09}) and subsequently the Zwicky Transient Facility (ZTF; \citealt{gra19,mas19}) based on a color selected sample from Pan-STARRS data release 1. The PTF used the Palomar 48$^{\prime\prime}$ Samuel Oschin Schmidt telescope to image up to $\approx2000$\,deg$^2$ of the sky per night to a depth of R$_{\rm mould} \approx20.6$\,mag or $g' \approx21.3$\,mag. PTF was succeeded by the Zwicky Transient Facility which started science operation in March 2018 using the same telescope but a new camera with a field-of-view of $47$\,deg$^2$. Here we report the discovery of a new thermonuclear supernova double detonation progenitor system consisting of an sdB with a WD companion: PTF1\,J223857.11+743015.1 (hereafter \ptfa) with orbital period of 76\,min showing similar properties to CD--30$^{\circ}$11223.


\section{Observations}

\subsection{Photometry}
As part of the Palomar Transient Factory (PTF), the Palomar 48-inch (P48) telescope imaged the sky every night. The reduction pipeline for PTF applies standard de-biasing, flat-fielding, and astrometric calibration to raw images \citep{lah14}. Relative photometry correction is applied and absolute photometric calibration to the few percent level is performed using a fit to SDSS fields observed in the same night \citep{ofe12}. The lightcurve of \ptfa\, has 144 epochs, with good photometry in the R$_{\rm mould}$ band with a typical uncertainty of 0.01-0.02\,mag. The majority of observations were conducted during the summer months June - August 2013 and 2014 and the cadence is highly irregular, ranging from a few minutes to years. The object was also observed as part of the Zwicky Transient Facility (ZTF) public survey \citep{gra19,bel19}. Image processing of ZTF data is described in full detail in \citet{mas19}. We extracted the light curve from ZTF data release 6 which consists of 34 observations in ZTF-$r$ taken randomly over $\approx$\,1.5 years between August 2018 and November 2019.

High-cadence observations were conducted using the Palomar 200-inch telescope with the high-speed photometer CHIMERA \citep{har16} which is a 2-band photometer which uses frame-transfer, electron-multiplying CCDs to achieve 15 ms dead time covering a $5\times5$ arcmin field of view.  Simultaneous optical imaging in two bands is enabled by a dichroic beam splitter centered at 567\,nm.  Data reduction was carried out with the ULTRACAM pipeline \citep{dhi07} customized for CHIMERA. All frames were bias-subtracted and flat-fielded. 1300 observations in $g^\prime$ and $r^\prime$ with a 5\,sec exposure time were obtained on 2017-07-26 and 2700 observations in $g^\prime$ and $i^\prime$ with a 4\,sec exposure time were obtained on 2017-12-14.

\begin{figure}
\begin{center}
\includegraphics[width=0.48\textwidth]{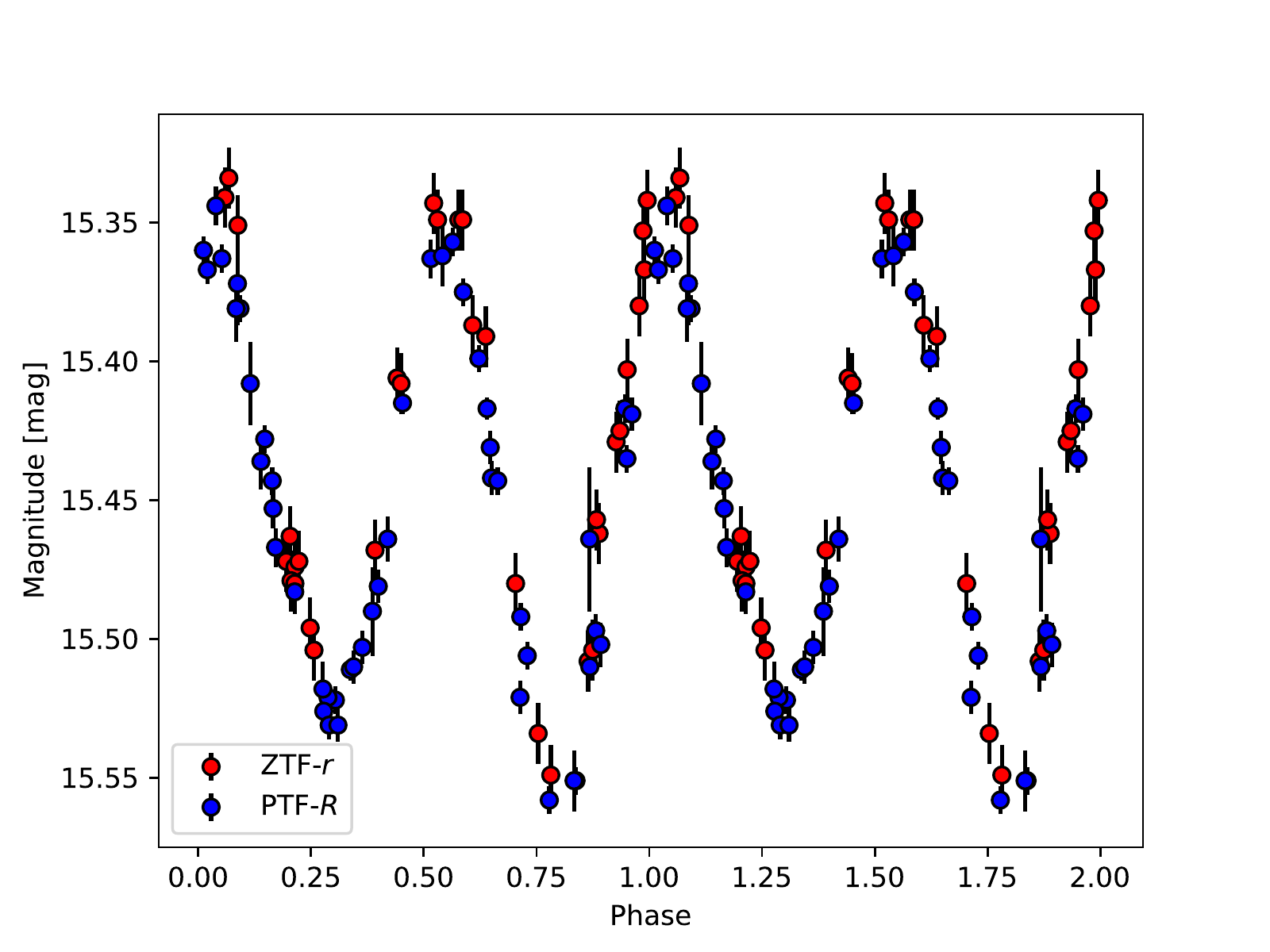}
\includegraphics[width=0.49\textwidth]{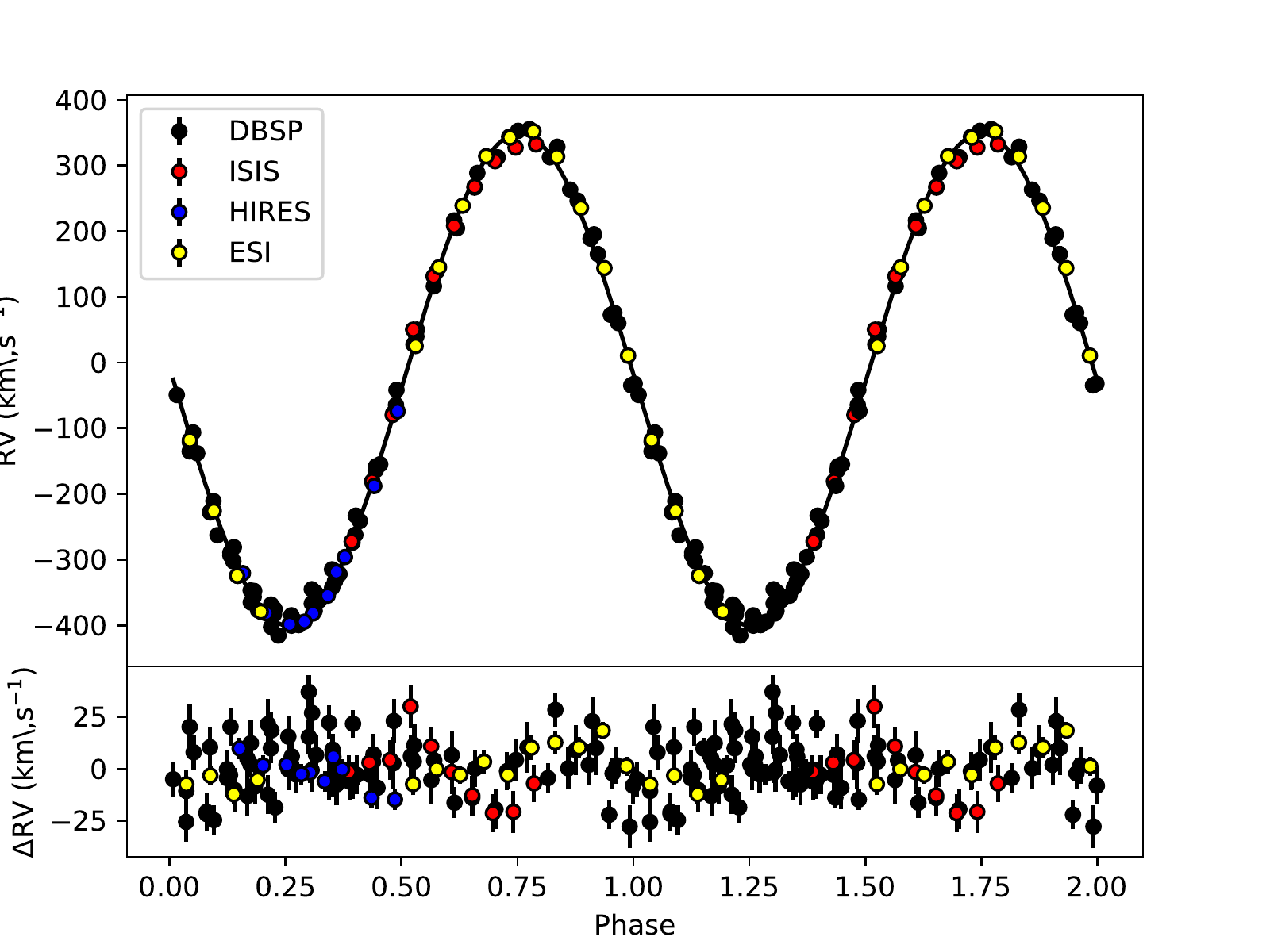}
\end{center}
\caption{{\bf Left panel:} Phase folded at \porb=$76.341750$\,min ZTF and PTF light curve for \ptfa. {\bf Right panel:} Radial velocity plotted against orbital phase for \ptfa. The RV data were phase folded with the orbital period and are plotted twice for better visualization. The residuals are plotted below.}
\label{fig:rv_curve1}
\end{figure}

\subsection{Spectroscopy}
Optical spectra were obtained with the Palomar 200-inch telescope and the Double-Beam Spectrograph (DBSP; \citealt{oke82}) using a low resolution mode ($R\sim1500$). 31 consecutive exposures were obtained on 2017-05-25 and 2017-05-29 and 15 consecutive exposures were obtained on 2017-05-25 using a 180\,sec exposure time. Each night an average bias and normalized flat-field frame was made out of 10 individual bias and 10 individual lamp flat-fields. To account for telescope flexure, an arc lamp was taken at the position of the target after each observing sequence. For the blue arm, FeAr and for the red arm, HeNeAr arc exposures were taken. Both arms of the spectrograph were reduced using a custom \texttt{PyRAF}-based pipeline \footnote{\url{https://github.com/ebellm/pyraf-dbsp}}\citep{bel16}. The pipeline performs standard image processing and spectral reduction procedures, including bias subtraction, flat-field correction, wavelength calibration, optimal spectral extraction, and flux calibration. 

Additionally \ptfa\, was also observed with the William Herschel Telescope (WHT) and the ISIS spectrograph \citep{car93} using a medium resolution mode (R600B grating, $R\approx2500$). 10 consecutive exposures with an exposure time of 180\,sec were obtained on 2017-07-26. 10 bias frames were obtained to construct an average bias frame and 10 individual lamp flat-fields were obtained to construct a normalized flat-field. CuNeAr arc exposures were taken before and after the observing sequence to correct for instrumental flexure. One dimensional spectra were extracted using optimal extraction and were subsequently wavelength and flux calibrated.

To obtain high-resolution spectra, \ptfa\, was observed with Keck/HIRES and Keck/ESI. We obtained 5 consecutive exposures with Keck/HIRES on 2017-08-14 and 2017-08-30 as well as 14 consecutive exposures with Keck/ESI on 2018-07-20. ThAr arc exposures were taken at the beginning of the night. The spectra were reduced using the \texttt{MAKEE}\footnote{\url{https://sites.astro.caltech.edu/~tb/makee/}} pipeline following the standard procedure: bias subtraction, flat fielding, sky subtraction, order extraction, and wavelength calibration. 

\begin{figure*}
\begin{center}
\includegraphics[width=0.48\textwidth]{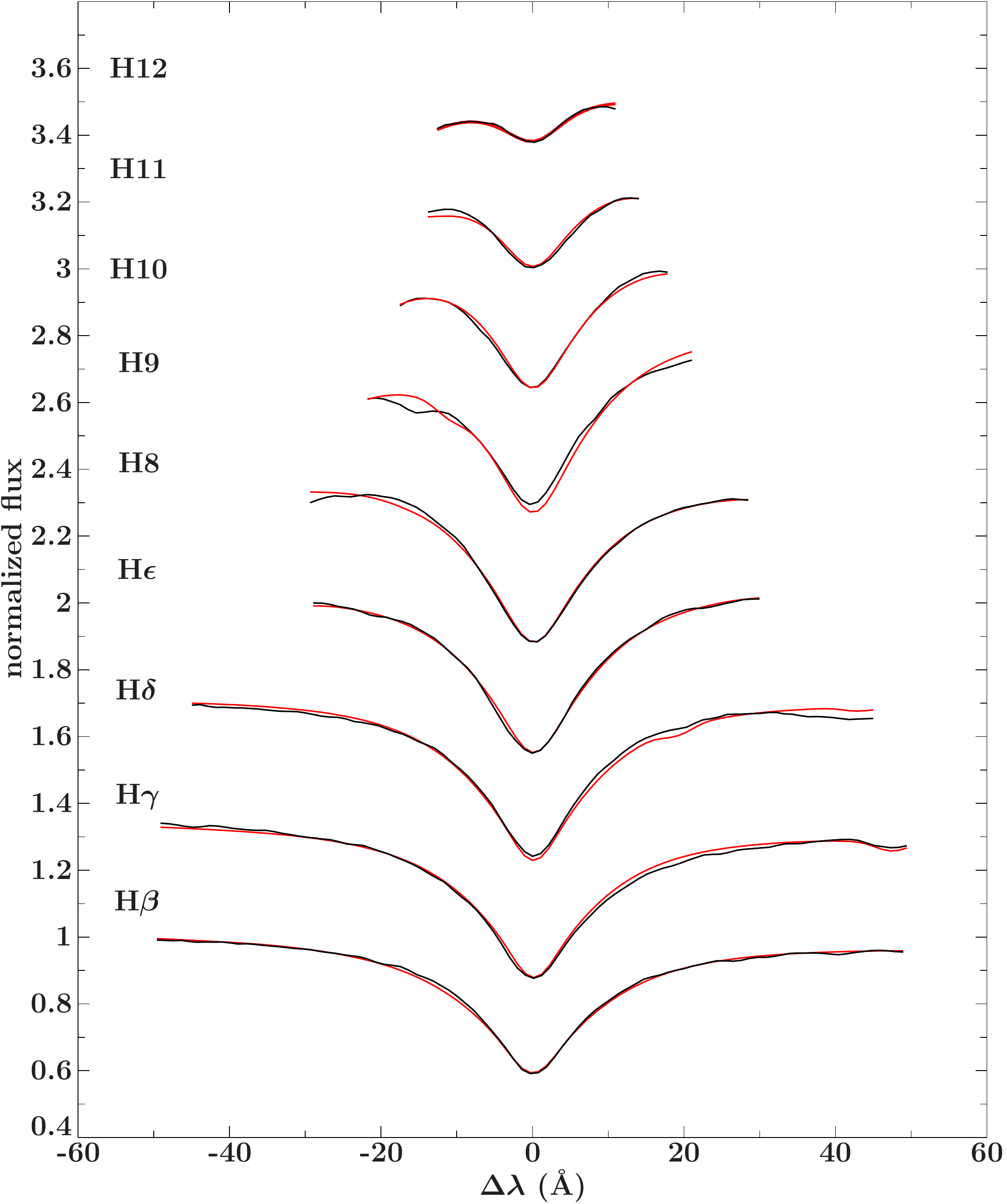}
\includegraphics[width=0.48\textwidth]{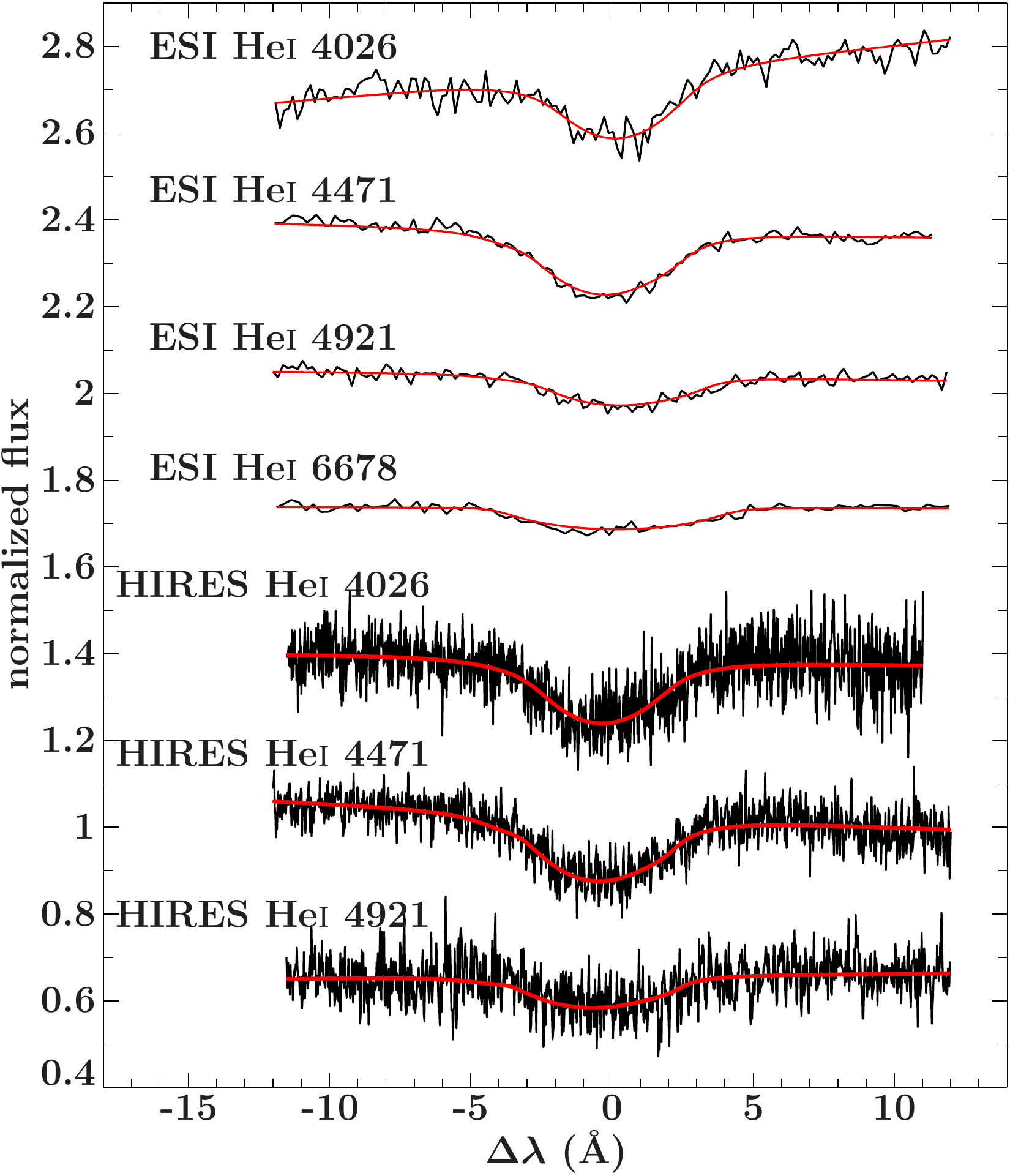} 
\caption{{\bf Left panel:} Fit of synthetic LTE models to the hydrogen Balmer lines of a coadded DBSP spectrum. The normalized fluxes of the single lines are shifted for better visualisation. {\bf Right panel:} Fits of \vrot\ to the helium lines seen in the HIRES and ESI spectra. The atmospheric parameters were fixed to the values derived from the WHT and DBSP spectra.}
\end{center}
\label{fig:rotation}
\end{figure*}

\section{Orbital and atmospheric parameters and light curve fitting}\label{orb_atm_pars}
As evident in Fig.\,\ref{fig:rv_curve1}\, \ptfa\ shows strong periodic ellipsoidal variability in its light curve at \porb\ = 76.341750(1)\,min. This variability is caused by the tidal deformation of the sdB primary under the influence of the gravitational force of the companion. We use the PTF and the ZTF lightcurve with its multi-year baseline and the Chimera light curves to derive the orbital period of the systems. The analysis was done with the \texttt{Gatspy} module for time series analysis which uses the Lomb-Scargle periodogram\footnote{http://dx.doi.org/10.5281/zenodo.14833} \citep{van15}. The error was derived from bootstrapping.

Radial velocities were measured by fitting Gaussians, Lorentzians, and polynomials to the hydrogen and helium lines to cover continuum, line, and line core of the individual lines using the \texttt{FITSB2} routine \citep{nap04a}. The procedure is described in full detail in \citet{gei11a}. We fitted the wavelength shifts compared to the rest wavelengths using a $\chi^2$-minimization. Assuming circular orbits, a sine curve was fitted to the folded radial velocity (RV) data points (Fig.\,\ref{fig:rv_curve1}).

Atmospheric parameters such as effective temperature, \teff, surface gravity, \logg, helium abundance, $\log{y}=\log\frac{n(He)}{n(H)}$, and projected rotational velocity, \vrot, were determined by fitting the rest-wavelength corrected average DBSP, ISIS and HIRES spectra with metal-line-blanketed LTE model spectra \citep{heb00}. \teff\ and \logg\ were derived from the Balmer and helium lines from the ISIS and DBSP spectra whereas $\log{y}$ and \vrot\ were measured with the HIRES spectra. High-resolution echelle spectra are not well suited to measure \teff\ and \logg\ because the broad hydrogen absorption lines span several individual echelle orders and merging of the echelle spectra could introduce systematic errors. The full procedure is described in detail in \citet{kup17,kup17a}. \ptfa\ shows typical \teff, \logg, and $\log{y}$ and \vrot=185$\pm$5\,\kms. The rotational velocity is consistent with a tidally locked sdOB star (see Sec.\,\ref{sec:systemparam}). Figure 2 shows the main Balmer and helium lines with the best fit to the data. Table\,\ref{tab:system} summarizes the atmospheric and orbital parameters.



\begin{table*}[t]
\centering
\caption{Overview of the measured and derived parameters for \ptfa}
\begin{tabular}{lll}
\hline\hline
Right ascension & RA [hrs]  & 22:38:57.11  \\
Declination  &  Dec $[^\circ]$  & +74:30:15.1 \\
Magnitude$^b$ & $g$ [mag] & 15.244$\pm$0.023 \\
Parallax$^a$  & $\varpi$    [mas] & $1.0001\pm0.0225$   \\ 
Distance      &  $d$ [kpc] & $1.00\pm0.03$    \\
Absolute Magnitude   &  \multirow{2}{*}{$M_{\rm g}$ [mag]}   & \multirow{2}{*}{$4.40\pm0.20$}   \\
(reddening corrected) &   &  \\
Proper motion$^a$ (RA) &   $\mu_\alpha$cos$(\delta)$  [mas\,yr$^{-1}$]   &  $0.344\pm0.056$   \\
Proper motion$^a$ (Dec) &   $\mu_\delta$  [mas\,yr$^{-1}$]   &  $-1.833\pm0.051$   \\
\hline
\multicolumn{3}{l}{\bf{Atmospheric parameters of the sdB}}     \\ 
Effective temperature$^c$ & \teff\,[K] & 23\,600$\pm$400  \\
Surface gravity$^c$    & \logg  & 5.42$\pm$0.06   \\
Helium abundance$^d$  & $\log{y}$  & $-$2.11$\pm$0.03 \\
Projected rotational velocity$^d$  & \vrot\,[\kms] &  185$\pm$5  \\
\hline
\multicolumn{3}{l}{\bf{Orbital parameters}}   \\ 

&  $T_0$ [BMJD UTC]  &  57960.47584170(3)    \\
Orbital period & \porb\,[min]  & 76.341750(1)  \\
RV semi-amplitude & $K$ [\kms] & $378.0\pm3.7$     \\
System velocity & $\gamma$\,[\kms] & $-6.2\pm2.14$    \\ 
Binary mass function & $f_{\rm m}$ [\msol] &  0.0597$\pm$0.0020   \\
\hline
\multicolumn{3}{l}{\bf{Derived parameters}}      \\

Mass ratio  &  $q = \frac{M_{\rm WD}}{M_{\rm sdB}}$  & $0.528\pm0.020$
  \\
sdB mass &  $M_{\rm sdB}$ [\msol] & $0.383\pm0.028$  \\ 
sdB radius & $R_{\rm sdB}$ [R$_{\odot}$] & $0.190\pm0.003$    \\ 
WD mass &  $M_{\rm WD}$ [\msol] &  $0.725\pm0.026$    \\
WD radius &  $R_{\rm WD}$ [\msol] & $0.0109^{+0.0002}_{-0.0003}$   \\
WD blackbody temperature &  \teff\,[K] & $26,800\pm4600$ \\
Orbital inclination & $i$\,[$^\circ$] & $88.4^{+1.6}_{-3.3}$   \\
Separation  & $a$ [R$_{\odot}$]   & $0.615\pm0.010$   \\
Roche filling factor  &  $R_{\rm sdB}$/$R_{\rm Roche lobe}$  &  $0.951\pm0.010$   \\
\hline
\end{tabular}
\begin{flushleft}
$^a$ from Gaia eDR3 \citep{gai16,gai21}\\
$^b$ from PanSTARRS DR1 \citep{cham16}\\
$^c$ adopted from from DBSP and ISIS\\
$^d$ adopted from ESI and HIRES
\label{tab:system}
\end{flushleft}
\end{table*}

\begin{figure*}
\begin{center}
\includegraphics[width=0.48\textwidth]{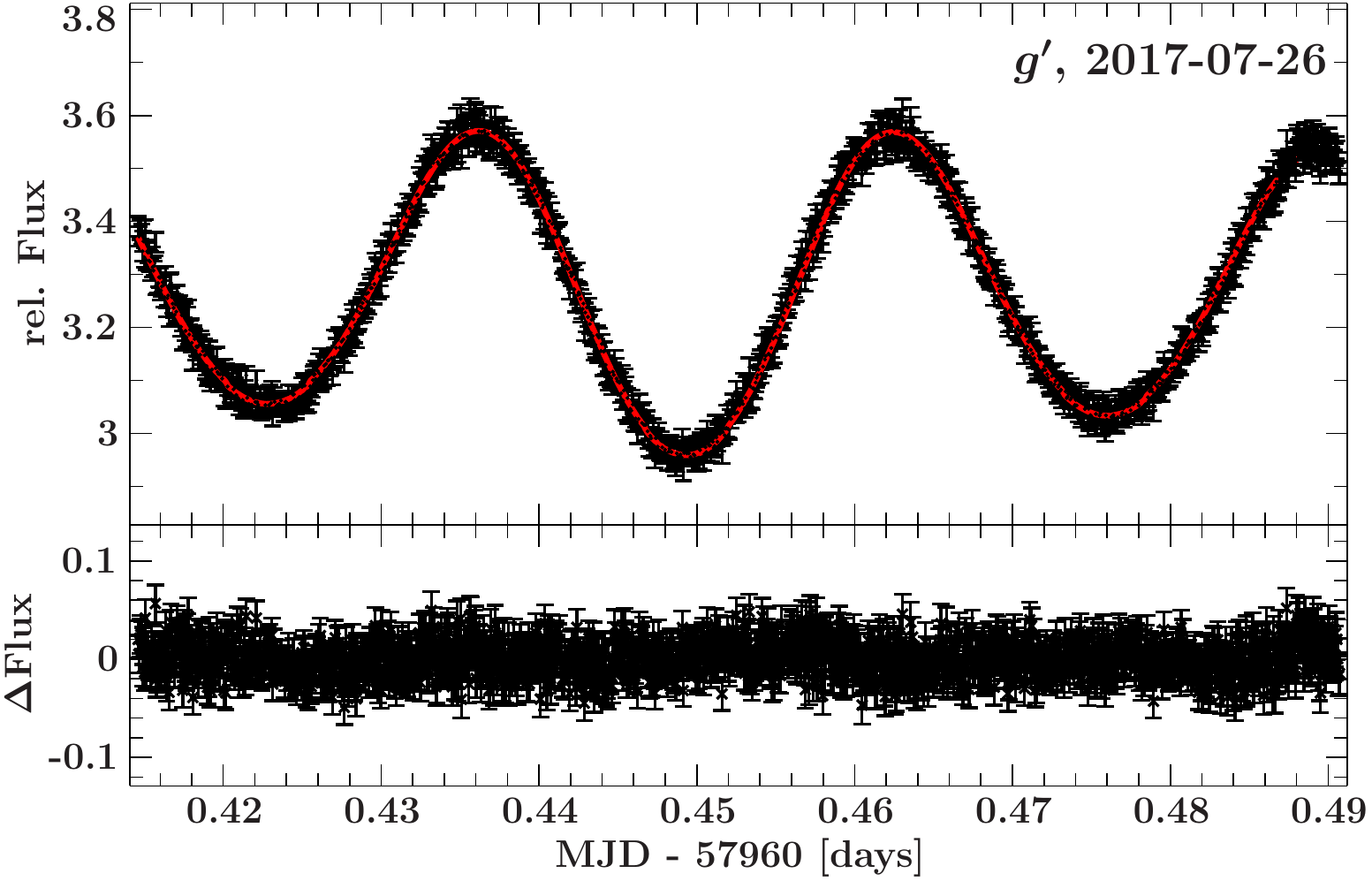}
\includegraphics[width=0.48\textwidth]{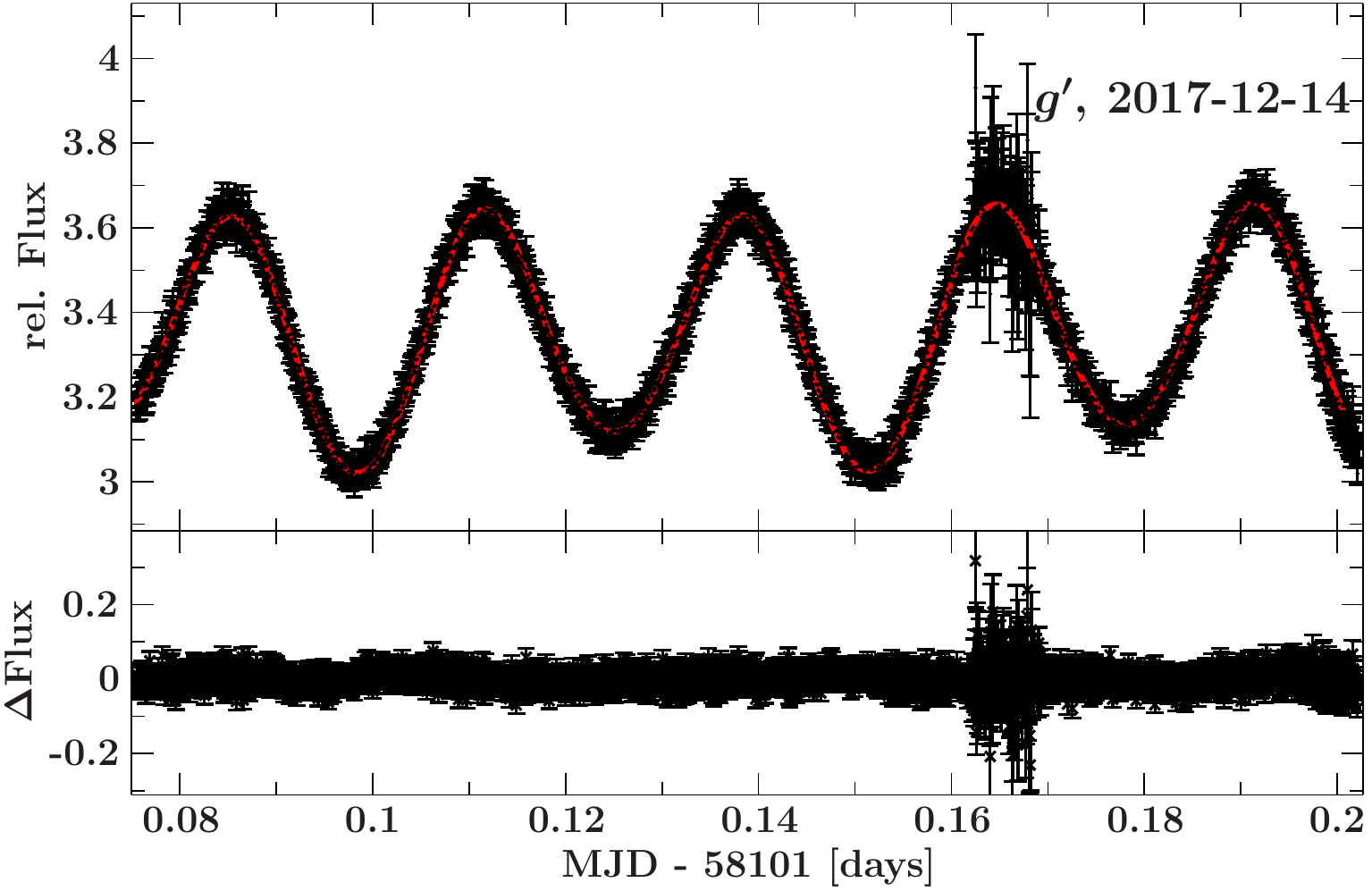}
\includegraphics[width=0.48\textwidth]{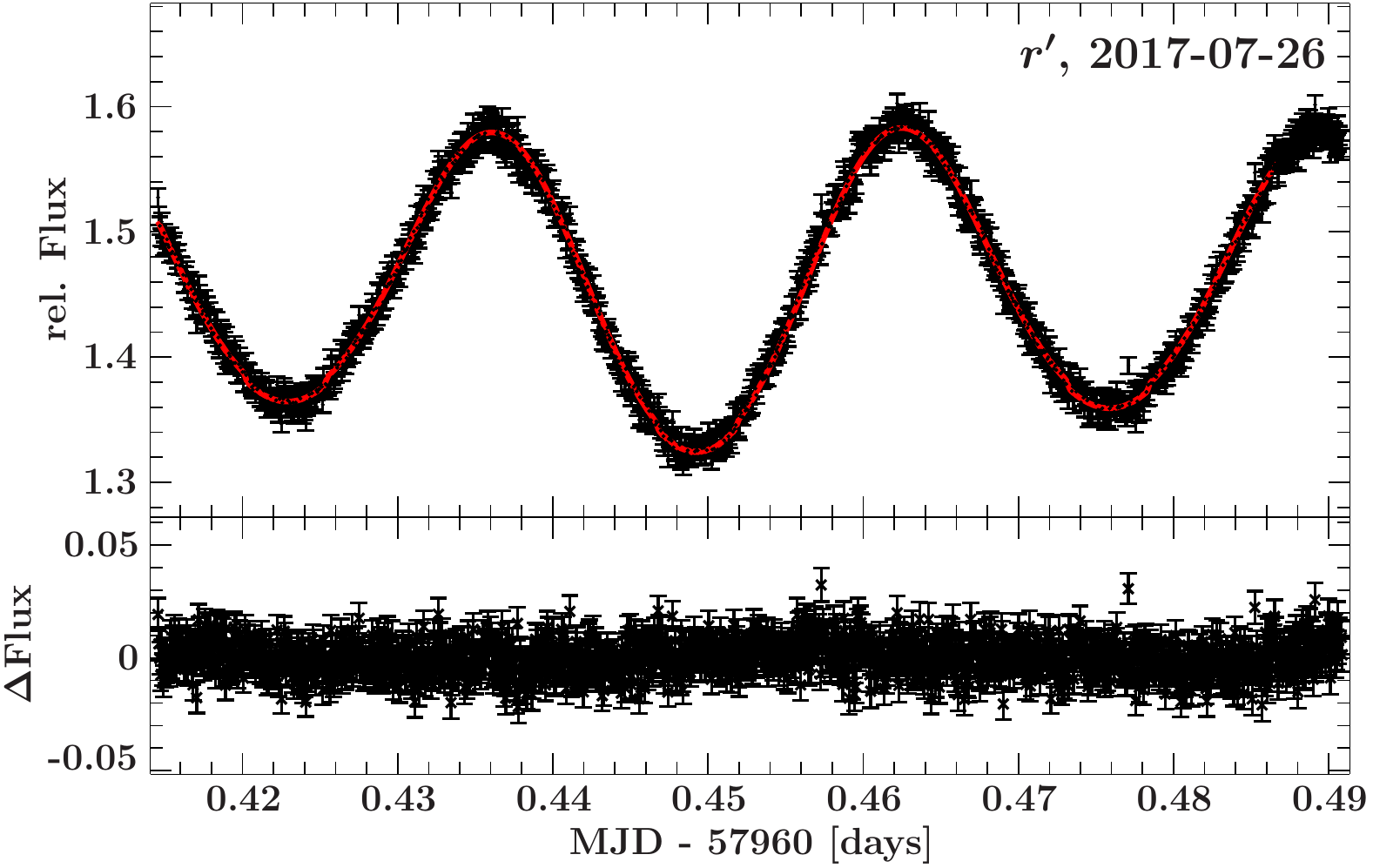}
\includegraphics[width=0.48\textwidth]{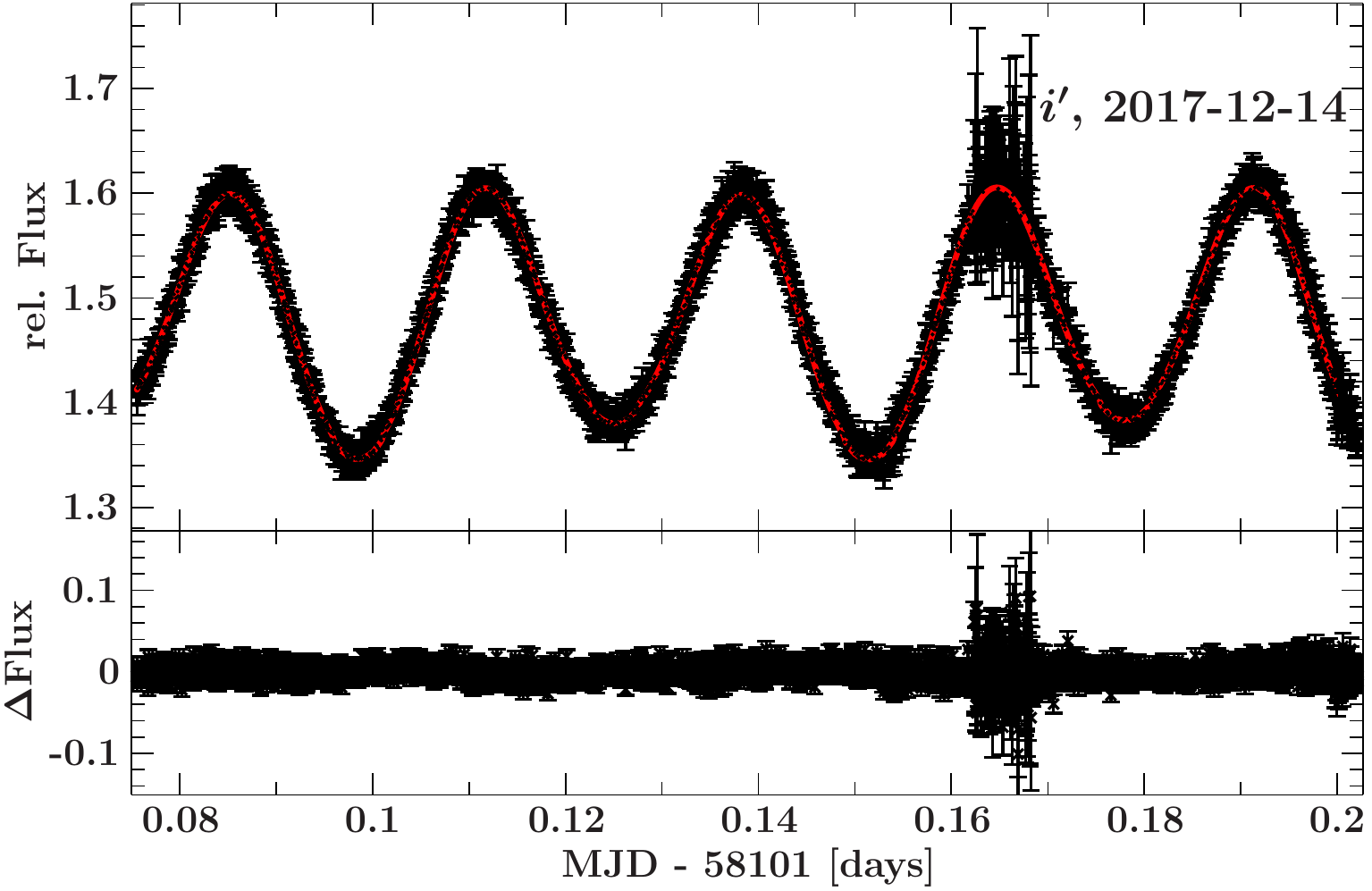}
\includegraphics[width=0.33\textwidth]{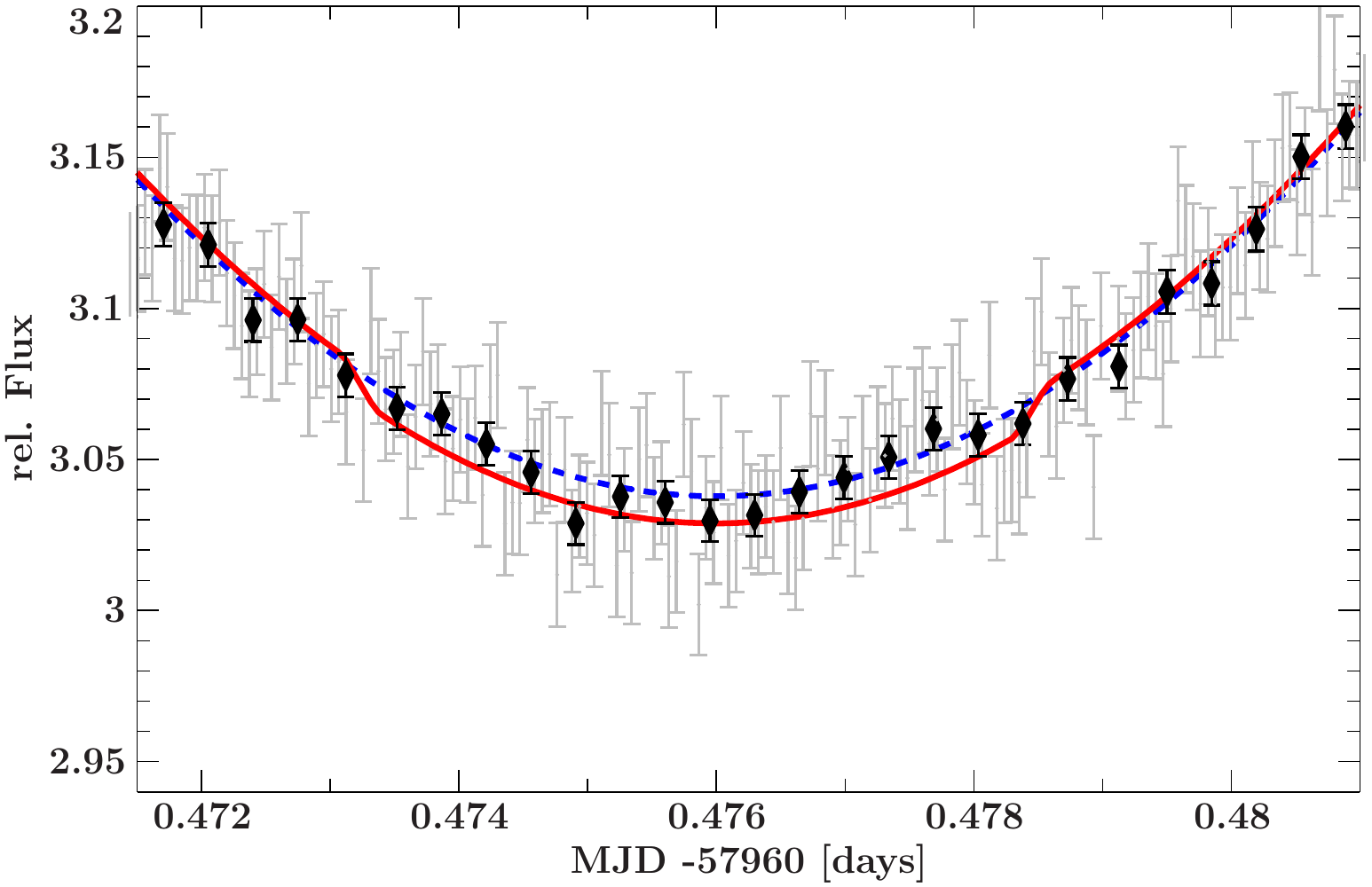}
\includegraphics[width=0.33\textwidth]{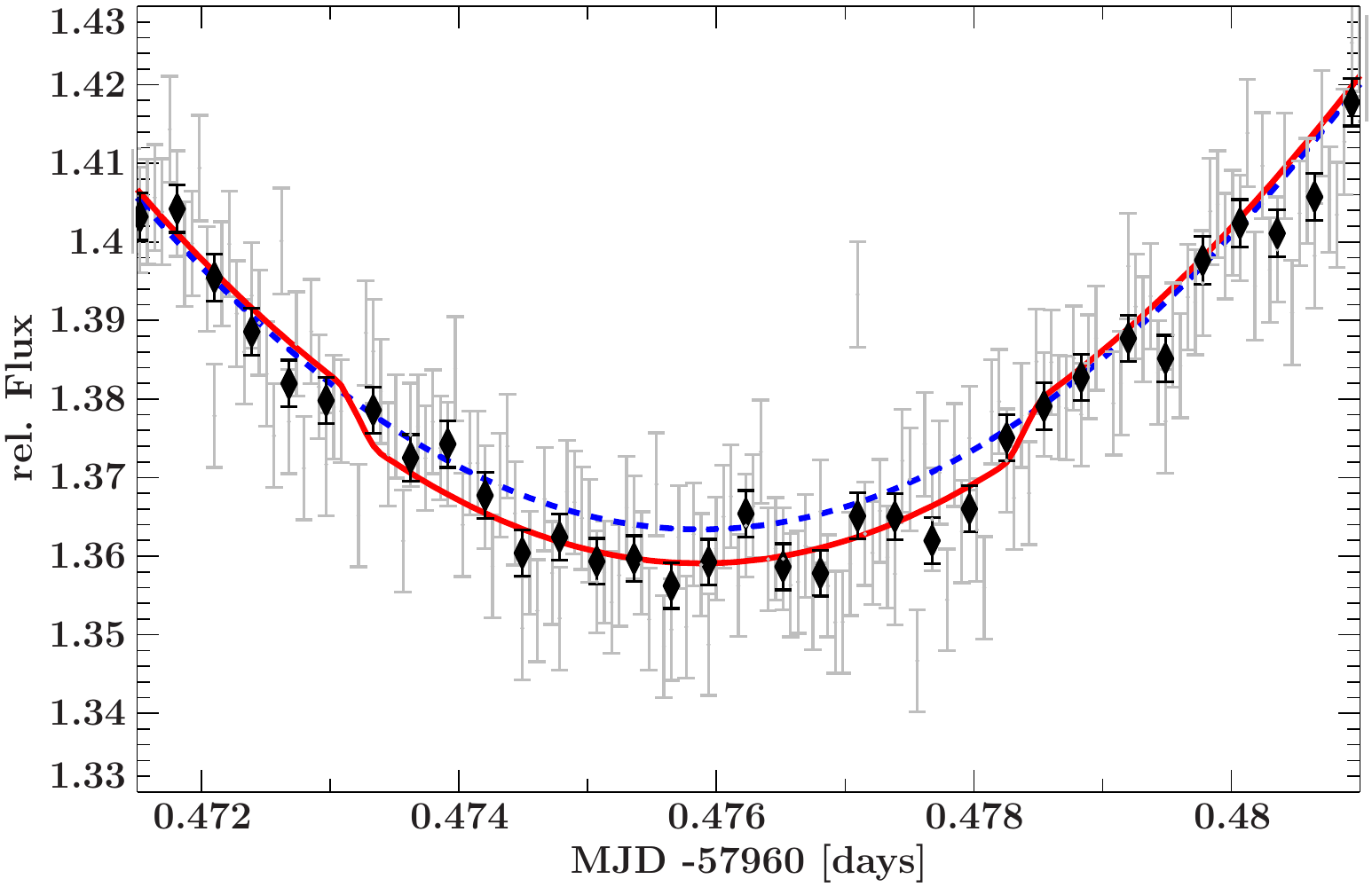}

\end{center}
\caption{Chimera light curves un-binned (grey) and binned (black) shown together with the \texttt{LCURVE} fits (red) observed optical SDSS bandpasses. The lower two panels show the region when the WD is being eclipsed by the sdB. The blue solid curve marks the same model without eclipses of the WD. The lower panels show the region when the white dwarf is being eclipsed. {\bf Lower left panel:} $g^\prime$ light curve, {\bf Lower right panel:} $r^\prime$ light curve }
\label{fig:light_fits_2130}
\end{figure*}

To model the lightcurves obtained with CHIMERA we used the \texttt{LCURVE} code \citep{cop10}. We use a Roche geometry, and the free parameters in our fit are: the phase ($t_0$), the scaled radii ($r_{1,2}$), the mass ratio $q$, the inclination $i$, secondary temperature $T_{\rm WD}$, and the velocity scale ($\mathrm[K+K_{\rm WD}]/\sin i$). We use a passband-dependent gravity-darkening law and use a gravity darkening value ($y_{g,r}$) from \cite{cla11} and find $\beta=0.425$ for $g^\prime$, $\beta=0.395$ for $r^\prime$, and $\beta=0.37$ for $i^\prime$. We assume an uncertainty of $0.03$ on the value and use a Gaussian prior. We use fixed limb darkening coefficients ($a_\mathrm{1}$, $a_\mathrm{2}$, $a_\mathrm{3}$, $a_\mathrm{4}$) taken from \cite{cla11}. We use $a_\mathrm{1}=0.82, a_\mathrm{2}=-0.65, a_\mathrm{3}=0.55$, and $a_\mathrm{4}=-0.19$ for $g^\prime$, $a_\mathrm{1}=0.81, a_\mathrm{2}=-0.89, a_\mathrm{3}=0.79$, and $a_\mathrm{4}=-0.27$ for $r^\prime$, and $a_\mathrm{1}=0.78, a_\mathrm{2}=-1.01, a_\mathrm{3}=0.91$, and $a_\mathrm{4}=-0.31$ for $i^\prime$. We also model the relativistic beaming ($F$) as in \cite{blo11}. We calculate the beaming parameters by assuming a blackbody spectrum and using the effective wavelength of the $g^\prime$, $r^\prime$, and $i^\prime$ filters. We find $F=1.80$ for $g^\prime$, $F=1.57$ for $r^\prime$, and $F=1.46$ for $i^\prime$. The full approach is also described in \citet{kup17,kup17a,kup20,kup20a} and \citet{rat19}. In addition, we add a 2nd order polynomial to correct for any long timescale trends which are the result of a changing airmass over the course of the observations. The best value of $\chi^2$ for this model was 1350 for 1300 data points for the g-band light curve which includes also a weak eclipse of the hot WD. Although the eclipse is weak ($\leq1$\,\%; Fig. 3), the $\chi^2$ for the non-eclipsing solution is 1400 which is statistically significantly worse compared to the solution with the weak eclipse. We use the MCMC sampler {\sc emcee} \citep{for13} to determine the best-fit values and uncertainty on the parameters. Figure 3 shows the Chimera light curves with the best fitted model. The lower panels are zoomed in around the region when the WD is being eclipsed.

\section{Results}
\subsection{System parameters}\label{sec:systemparam}
Although, \ptfa\, is a single-lined binary we can derive system parameters using the combined results from the light curve analysis with results from the spectroscopic fitting. Parameters derived in this way by a simultaneous fit to the Chimera light curves are summarized in Table \ref{tab:system}. The given errors are all $95\,\%$ confidence limits.

We find that \ptfa\, consists of a low mass sdB with a high-mass WD companion. We derive a mass ratio $q = M_{\rm sdB}/M_{\rm WD}=0.528\pm0.020$, a mass for the sdB  $M_{\rm sdB}=0.383\pm0.028$\,\msol, and a WD companion mass $M_{\rm WD}=0.725\pm0.026$\,\msol. \ptfa\, is found to be eclipsing at an inclination angle of $i = 88.4^{+1.6}_{-3.3}\,^\circ$ which allows us to measure the radius and the black-body temperature of the WD companion. We determine a black-body temperature of $26,800\pm4600$\,K for the WD and a radius of $R_{\rm WD}=0.0109^{+0.0002}_{-0.0003}$\,\rsol. The radius was found to be $<5\%$\, above the zero-temperature value and is fully consistent with predictions from \citet{rom19} for carbon-oxygen core white dwarfs.

\citet{zah77} predicted that the sdBs in close sdB binaries with orbital periods below $\approx0.3$\,days should be synchronized to the orbit. More recently, \citet{pre18} found that only the most compact sdB binaries should be synchronized. From the system parameters we find that the sdB would have a projected rotational velocity \vrot$=181\pm6$\,\kms\, if synchronized to the orbit. The measured \vrot$=185\pm5$\,\kms\, is consistent with a synchronized orbit.


We calculate the absolute magnitude ($M_{\rm g}$) of \ptfa\, using the visual PanSTARRS g-band magnitude g=15.244$\pm$0.023\,mag and the parallax from Gaia eDR3 \citep{gai16,gai21}. Because the object is located near the Galactic Plane, significant reddening can occur. \citet{gre19} present updated 3D extinction maps based on Gaia parallaxes and stellar photometry from Pan-STARRS 1 and 2MASS\footnote{http://argonaut.skymaps.info/} and find towards the direction of \ptfa\, an extinction of $E(g-r)=0.24\pm0.03$ at a distance of $1.00$\,kpc; this results in a total extinction in the g-band of $A_{\rm g}=0.84\pm0.11$\,mag, and with the corrected magnitude, we find an absolute magnitude of $M_{\rm g}=4.40\pm0.20$\,mag consistent with a hot subdwarf star \citep{gei19}.

\subsection{Comparison with Gaia parallax}\label{sec:gaia}
To test whether our derived system parameters are consistent with the parallax provided by Gaia eDR3, we compared the measured parameters from the light curve fit to the predictions using the Gaia parallax. The approach follows a similar strategy as described in \citet{rat19} and \citet{kup20}. Using the absolute magnitude $M_{\rm g}=4.40\pm0.20$\,mag, we find a luminosity of $L=11.5\pm3.0$\,L$_\odot$ using a bolometric correction $BC_{\mathrm{g}}=-2.30$\,mag derived for our stellar parameters from the {\tt MESA} Isochrones \& Stellar Tracks (MIST; \citealt{dot16, cho16, pax11, pax13, pax15, pax18}). Using the Stefan-Boltzmann law applied to a black body ($L = 4 \sigma \pi R_{\rm{sdB}}^{2}T_{\rm{eff}}^4$), we can solve for the radius of the sdBs, and combined with $R_{\rm{sdB}}^{2}=GM_{\rm{sdB}}/g$, we can solve for mass of the sdBs:
\begin{equation}
M_{\rm{sdB}} = \frac{L_{\rm{sdB}}10^{\log(g)}}{4\pi\sigma GT_{\rm{eff}}^4}
\end{equation}
Using these equations we find $M_{\rm sdB} = 0.39\pm0.10$\,\msol\, and $R_{\rm sdB} = 0.17\pm0.03$\,\rsol. Although the error bars are rather large, this result is in agreement with the results from the light curve and spectroscopic fits.

 

\subsection{Kinematics of the binary systems}
We find that \ptfa\, has evolved from a $\approx$2\,\msol\ star (see Sect.\,\ref{sec:future}), and we expect the system is a member of a young stellar population. 
 Using the proper motion from Gaia eDR3 \citep{gai16,gai18,gai21}, the distance and the systemic velocities (see Tab.\,\ref{tab:system}) we calculate the Galactic motion for \ptfa.  

We employed the approach described in \citet{ode92} and \citet{pau06}. As in \citet{kup20}, we use the Galactic potential of \citet{all91} as revised by \citet{irr13}. The orbit was integrated from the present to 3 Gyr into the past. We find that the binary moves within a height of 200\,parsec of the Galactic equator and with very little eccentricity between 9 and 10\,kpc from the Galactic center. From the Galactic orbit we conclude that \ptfa\, is a member of the Galactic thin disk population consistent with being member of a young stellar population.

\begin{figure}
    \centering
    \includegraphics[width=0.9\textwidth]{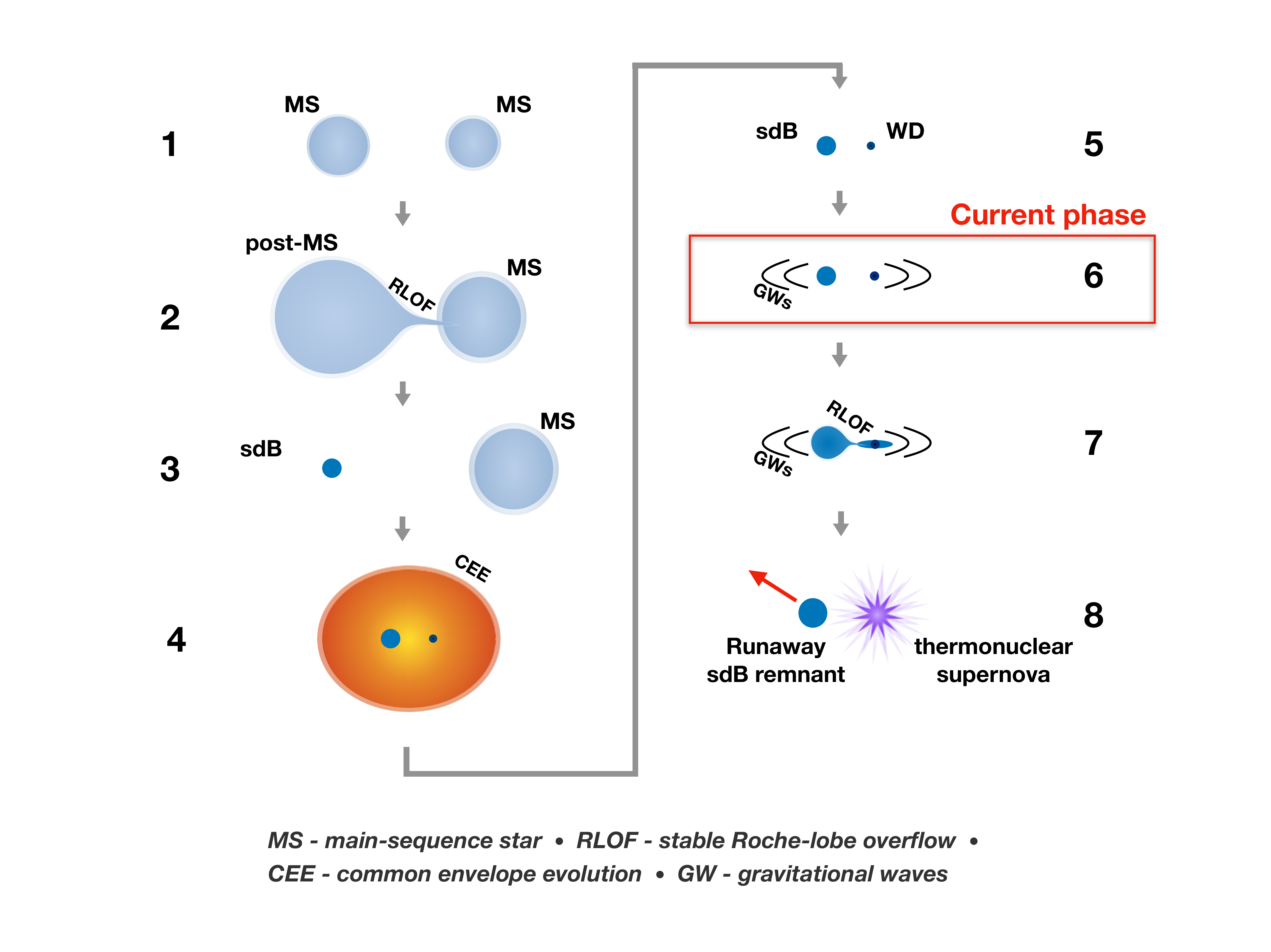}
    \caption{Visualization of the proposed evolutionary pathway for \ptfa. The red box marks the current evolutionary phase. Each evolutionary phase is numbered according to their order in the evolution and the direction of the sequence is marked with arrows.}
    \label{fig:sdb_evolv}
\end{figure}

\begin{figure}
    \centering
    \includegraphics[width=0.54\textwidth]{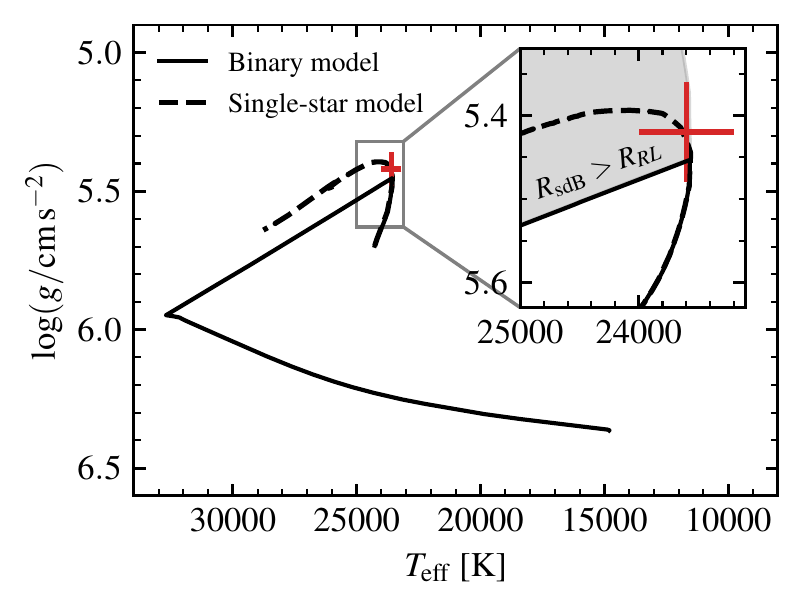}
    \includegraphics[width=0.44\textwidth]{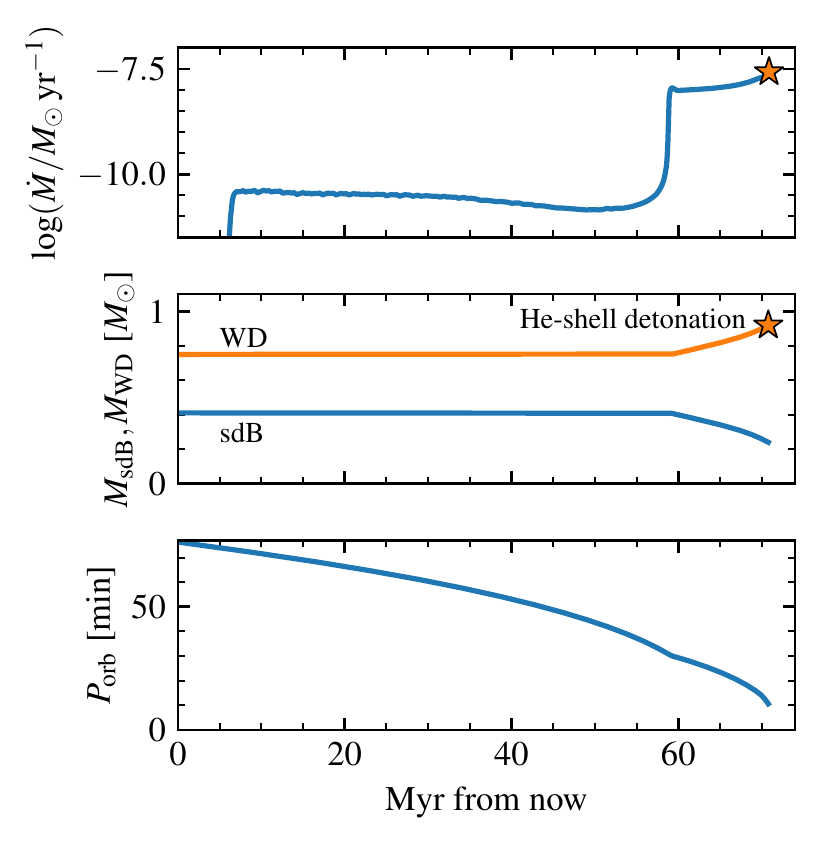} 
    \caption{{\bf Left panel:} Predicted evolution based on the {\tt MESA} model for the PTF1 J2238+7430 system. The current observed \logg\ and \teff\ and error bars for the system are shown in red. The dashed curve shows the evolution the star would follow in isolation, while the solid curve shows the trajectory it follows due to encountering the Roche limit, depicted by the gray shaded region {\bf in the inset}. {\bf Right panel:} Future evolution of the system until the helium ignites.}
    \label{fig:kiel_mesa}
\end{figure}

\section{Predicted Evolution of the binary system}\label{sec:evolution}

\subsection{Formation of the sdB + WD system}
\citet{rui10} found that the dominant way to form compact double carbon-oxygen core WDs is through stable mass transfer which forms the sdB followed by a phase of unstable mass transfer which forms the white dwarf companion. They present a specific example which starts with a $2.88$\,\msol\, and $2.45$\,\msol\, binary pair. In \ptfa\, weak eclipses of the WD companion imply a blackbody temperature of $26\,800\pm4600$\,K. From the blackbody temperature we can estimate the cooling age and find a cooling time of $\approx$25\,million years, significantly shorter than the predicted current age of the sdB of $\approx$170\,million years (see Sec.\,\ref{sec:future}). Therefore, we predict that the sdB was formed first, and we propose the following evolutionary scenario (illustrated in Fig.~\ref{fig:sdb_evolv}) for \ptfa\,which explains all observational properties and is similar to the scenario discussed in \citet{rui10}. 

The system started as a $\approx 2$\,\msol\, main sequence star (see Sect.\,\ref{sec:future}) which will become the sdB, and a slightly lower mass companion with an orbital period of a few weeks. The sdB progenitor evolves first and starts stable mass transfer to the companion star.  At the end of that phase the sdB has formed with the observed mass of $\approx0.4$\,\msol\, and the orbital periods has substantially widened consistent with the first stable RLOF channel described in \citet{han02,han03}. The companion star has accreted $\approx1.7$\,\msol\ of material from the sdB progenitor and turned into a $\approx$3.5--4\,\msol\, star which will then evolve off the main sequence and overflow its Roche Lobe while the sdB star is still burning helium. Due to the large mass ratio at this point, mass transfer will be unstable and initiate a common envelope. The CE phase could happen either during the RGB or AGB phase of the secondary depending on the binary separation at that point. In either case it would leave a compact binary with a massive WD and an sdB at an orbital period of $\approx86$\,minutes. The observed high WD mass of $0.725\pm0.026$\,\msol\, is consistent with the evolution from an intermediate mass main sequence star \citep{cum18}. The final phase of unstable mass transfer happened $\approx25$\, million years ago, after which the WD cooled to its currently observed temperature while gravitational wave radiation decreased the orbital period to the currently observed period of $76$\,minutes. As also discussed in \citet{rui10}, there could exist a substantial fraction of compact sdB+WD binaries where the sdB was formed first through stable mass transfer.


\subsection{Future evolution}\label{sec:future}
To understand the future evolution of the system we employed {\tt MESA} version 12115 \citep{pax11,pax13,pax15,pax18,pax19}. \cite{bau21} use {\tt MESA} models to show that sdB stars with mass $M \lesssim 0.47$\,\msol\ can descend from either lower-mass main sequence progenitors that ignite central He burning via an off-center degenerate He flash ($M_{\rm ZAMS} \lesssim 2.3$\,\msol), or they can descend from higher-mass main sequence progenitors that ignite He at the center under non-degenerate conditions ($M_{\rm ZAMS} \gtrsim 2.3$\,\msol).%
\footnote{The precise value of the progenitor $M_{\rm ZAMS}$ for which He ignition conditions change depends somewhat on metallicity and overshoot \citep{ost21}, but generally lies between about 2.0 and 2.3\,\msol.}
They show that these scenarios lead to different H envelope structures that influence the subsequent radius evolution of the sdB star, with stars descended from higher mass progenitors having more compact envelopes and correspondingly higher \logg\ values, as shown in the top panels of figure~5 in \cite{bau21}.
The measured \logg\ for \ptfa\ requires a relatively extended envelope with a radius that requires that the sdB star descended from the lower-mass channel with a progenitor mass around 2\,\msol.
We find that our best matching {\tt MESA} model for the measured \logg\ and \teff\ of this system is a 0.41\,\msol\ sdB model descended from a 2.14\,\msol\ main sequence star that ignited the He core via a degenerate He-core flash. This model has a sharp transition from the He core to an H envelope with solar composition. When He ignites, we remove most of the envelope, leaving a thin H envelope layer of $10^{-3}$\,\msol\ so that the subsequent sdB evolution track matches the observed \logg\ and \teff\ of \ptfa.
Figure~\ref{fig:kiel_mesa} shows the \logg--\teff\ evolution of this {\tt MESA} model, where it approaches the current observed state of \ptfa\ after $\approx$170~Myr of evolution, and will encounter its Roche lobe and begin transferring mass soon after.

We model the future binary evolution of this system with a 0.75\,\msol\ WD companion using the {\tt MESA} binary capabilities. The WD model is constructed with a C/O core using the {\tt make\_co\_wd} test case from {\tt MESA}, rescaled to a mass of 0.75\,\msol, and cooled to the current observed temperature before initializing it into the {\tt MESA} binary model at the currently observed orbital period with the sdB model.
The sdB is currently observed at 95\,\% Roche Lobe filling and will continue to spiral in due to gravitational wave radiation. 
In our model the sdB will soon fill its Roche lobe and start to donate its hydrogen rich envelope in six million years at a low rate of $\lesssim 10^{-10}$\,\msol\,yr$^{-1}$ \citep[see][for a detailed oberview]{bau21}. 
Because of the large initial radius of the H envelope, mass transfer will proceed at this low rate for $\approx$50~Myr before the H envelope is exhausted and the He core is finally exposed at a much more compact radius.
While the sdB is still helium core burning $\approx$60 Myr from today, the sdB will begin to donate helium rich material onto the WD at the expected rate of {$\approx$}{1--3}$\times10^{-8}$\,\msol\,yr$^{-1}$, as shown in Figure~\ref{fig:kiel_mesa}.
A helium rich layer will slowly build up for $10$\,million years, reaching a critical mass of $0.17$\,\msol, after which the {\tt MESA} WD model experiences He ignition in the accreted envelope. At this point the binary has an orbital period of $\approx10$\,min. The sdB has been stripped down to a mass of $0.25$\,\msol, and the WD has a total mass of $0.92$\,\msol.

Our {\tt MESA} model predicts that at this point the accreting WD will experience a thermonuclear instability that will lead to a detonation that will likely destroy the WD in a thermonuclear supernova \citep{woo11,bau17}.
Our {\tt MESA} model for the WD accretor includes the NCO reaction chain as in \cite{bau17}, and this governs ignition in the accreted He envelope. Because this ignition mechanism is initiated by electron captures on $^{14}$N, it occurs at a density above $\rho = 10^6\,\rm g\,cm^{-3}$ where a detonation is likely to form \citep{woo94,woo11}.
The structure of our {\tt MESA} model at the point of detonation is very similar to the model for CD--30$^{\circ}$11223 in \cite{bau17}, which includes a more detailed discussion of detonation formation under these ignition conditions.
At the time of the thermonuclear supernova, the sdB remnant has an orbital velocity of $911$\,\kms\, and will be released as a hyper-runaway star exceeding the escape velocity of the Galaxy \citep{bau19,neu20,neu21, liu21}. Fig.\,\ref{fig:sdb_evolv} illustrates the evolutionary sequence proposed for \ptfa. 





\section{Supernova rate estimate}
Models of thermonuclear supernovae in WDs with thick ($\gtrsim 0.1$\,\msol) helium shells indicate that they will yield transients classified as peculiar Type I supernovae \citep{pol19,de19}. \ptfa\, together with CD-30$^\circ$11223 therefore mark a small sample of double detonation peculiar thermonuclear supernova progenitors. Using both systems we can estimate a lower limit of thermonuclear supernovae originating in compact hot subdwarf + WD binaries where the sdB donates helium rich material during helium core burning. Both systems will have an age of $\approx$500\,Myrs at the time of the helium shell detonation and are located within 1\,kpc. Because of their young age, we compare the rate of these double detonation progenitors to the supernova Ia rate as a function of star formation. Under the assumption that these systems typically have an age of $\approx$500\,Myrs at time of explosion we find a lower limit of double detonation explosions of $\frac{2}{500}$\,kpc$^{-2}$Myr$^{-1}$ from the two known systems. We can compare that to the local star formation rate of $10^{-3}$\,\msol kpc$^{-2}$yr$^{-1}$ which leads to a double detonation rate of $\approx4\times10^{-6}$\,yr$^{-1}$. \citet{sul06} found a supernova Ia rate of $3.9\pm0.7\times10^{-4}$\,SNe\,yr$^{-1}$  (\msol\,yr$^{-1}$)$^{-1}$ of star formation. With a Galactic star formation rate of $\approx$1\msol\,yr$^{-1}$, we find that the rate at which peculiar thermonuclear supernovae with thick $\approx0.15$\,\msol\ helium shells occur in star forming galaxies could be at least 1 \% of the type Ia supernova rate. This is in reasonable agreement with the presently observed low rate of thick helium shell detonations. We note that thermonuclear supernovae with thick helium layers are likely to produce a transient that would be classified as a peculiar SN Ia with lower luminosities and redder color compared to ordinary SN Ia \citep{pol19}.

\citet{de19} presented the discovery of peculiar Type I supernova consistent with a thick helium shell double detonation on a sub-Chandrasekhar-mass WD \citep{pol19,pol21}. However, one of the distinct differences is that the transient occurred in the outskirts of an elliptical galaxy which points to an old stellar population which is in disagreement with our observed systems which represent a young population. More recently, \citet{de20} presented a sample of calcium rich transients originating from double-detonations with helium shells. They find that the majority of transients are located in old stellar populations. However, \citet{de20} note that a small subsample (iPTF16hgs, SN2016hnk and SN 2019ofm) were found in star forming environments, suggesting that there is a small but likely non-zero contribution from young systems which could potentially be related to systems like CD-30$^\circ$11223 and \ptfa.

\section{Summary and Conclusion}
As part of our search for short period sdB binaries we discovered \ptfa\, using PTF and subsequently ZTF light curves. We find a period of \porb=76.34179(2)\,min. Follow-up observations confirmed the system as an sdB with $M_{\rm sdB}=0.383\pm0.028$\,\msol\, and a WD companion with $M_{\rm WD}=0.725\pm0.026$\,\msol. High-speed photometry observations with Chimera revealed a weak WD eclipse which allows us to measure the blackbody temperature and radius of the WD. We find a temperature of $26,800\pm4600$\,K and a radius of $R_{\rm WD}=0.0109^{+0.0002}_{-0.0003}$\,\rsol\, fully consistent with cooling models for carbon-oxygen core WDs. We find a cooling age of $\approx25$\,Myrs for the WD which is significantly shorter than our age estimate for the sdB which is $\approx$170\,Myrs. This can be explained by the sdB forming first through stable mass transfer, followed by the WD forming $\approx25$\,Myrs ago through a common envelope phase. This shows that evolutionary scenarios where the sdB is formed first through stable mass transfer must be considered for compact sdB binaries with WD companions. 

We employed \texttt{MESA} to calculate the future evolution of the system, finding that the sdB in \ptfa\, will start mass transfer of the hydrogen rich envelope in $\approx$6\,Myr. In $\approx$60\,Myr, after a phase of hydrogen and helium mass transfer, the WD will build up a helium layer of 0.17\,\msol\, leading to a total WD mass of 0.92\,\msol. Our models predict that at this point the WD will likely detonate in a peculiar thermonuclear supernova making \ptfa\, the second known progenitor for a supernova with a thick helium layer. Using both systems we estimate that at least 1\,\% of type Ia supernova originate from compact sdB+WD binaries in young populations of galaxies with similar star formation rates compared to the Milky Way. Although this is only a lower limit the estimate is broadly consistent with the low number of observed peculiar thermonuclear supernovae.

\acknowledgments

This research benefited from interactions that were funded by the Gordon and Betty Moore Foundation through grant GBMF5076. This work was supported by the National Science Foundation through grants PHY-1748958 and ACI- 1663688. TK would like to thank Ylva G\"{o}tberg for providing the template for Fig.\,\ref{fig:sdb_evolv}. TK acknowledges support from the National Science Foundation through grant AST \#2107982. DS was supported by the Deutsche Forschungsgemeinschaft (DFG) under grants HE 1356/70-1 and IR 190/1-1.

Observations were obtained with the Samuel Oschin Telescope at the Palomar Observatory as part of the PTF project, a scientific collaboration between the California Institute of Technology, Columbia University, Las Cumbres Observatory, the Lawrence Berkeley National Laboratory, the National Energy Research Scientific Computing Center, the University of Oxford, and the Weizmann Institute of Science.

Based on observations obtained with the Samuel Oschin 48-inch Telescope at the Palomar Observatory as part of the Zwicky Transient Facility project. ZTF is supported by the National Science Foundation under Grant No. AST-1440341 and a collaboration including Caltech, IPAC, the Weizmann Institute for Science, the Oskar Klein Center at Stockholm University, the University of Maryland, the University of Washington, Deutsches Elektronen-Synchrotron and Humboldt University, Los Alamos National Laboratories, the TANGO Consortium of Taiwan, the University of Wisconsin at Milwaukee, and Lawrence Berkeley National Laboratories. Operations are conducted by COO, IPAC, and UW.

Some of the data presented herein were obtained at the W.M. Keck Observatory, which is operated as a scientific partnership among the California Institute of Technology, the University of California and the National Aeronautics and Space Administration. The Observatory was made possible by the generous financial support of the W.M. Keck Foundation. The authors wish to recognize and acknowledge the very significant cultural role and reverence that the summit of Mauna Kea has always had within the indigenous Hawaiian community. We are most fortunate to have the opportunity to conduct observations from this mountain. 

Some results presented in this paper are based on observations made with the WHT operated on the island of La Palma by the Isaac Newton Group in the Spanish Observatorio del Roque de los Muchachos of the Institutio de Astrofisica de Canarias.

This work has made use of data from the European Space Agency (ESA) mission
{\it Gaia} (\url{https://www.cosmos.esa.int/gaia}), processed by the {\it Gaia}
Data Processing and Analysis Consortium (DPAC,
\url{https://www.cosmos.esa.int/web/gaia/dpac/consortium}). Funding for the DPAC
has been provided by national institutions, in particular the institutions
participating in the {\it Gaia} Multilateral Agreement.


\facilities{PO:1.2m (PTF), PO:1.2m (ZTF), Hale (DBSP), ING:Herschel (ISIS), Keck:I (HIRES), Keck:II (ESI),  Hale (Chimera)}

\software{\texttt{Gatspy} \citep{van15, van15a}, \texttt{FITSB2} \citep{nap04a}, \texttt{LCURVE} \citep{cop10}, \texttt{emcee} \citep{for13}, \texttt{MESA} \citep{pax11,pax13,pax15,pax18,pax19}, \texttt{Matplotlib} \citep{hun07}, \texttt{Astropy} \citep{astpy13, astpy18}, \texttt{Numpy} \citep{numpy}, \texttt{ISIS} \citep{2000ASPC..216..591H}, MAKEE (\url{https://sites.astro.caltech.edu/~tb/makee/}) }

\bibliography{refs,refs_1508}{}
\bibliographystyle{aasjournal}

\end{document}